\documentclass[aps,twocolumn,showpacs,preprintnumbers,amsmath,amssymb,nofootinbib]{revtex4}
\usepackage{amssymb}
\usepackage{epsfig}
\usepackage{graphicx}

\usepackage{graphicx}
\usepackage{dcolumn}
\usepackage{bm}

\begin{document}

\preprint{APS/123-QED}

\title{Some of semileptonic and nonleptonic decays of $B_c$ meson
in a Bethe-Salpeter relativistic quark model}

\author{Chao-Hsi Chang}
\email{zhangzx@itp.ac.cn} \affiliation{China Center of Advanced Science and Technology (World Laboratory), Beijing 100190, China;\\
State Key Laboratory of Theoretical Physics, Institute of Theoretical Physics, Chinese Academy of Sciences, Beijing 100190, China;}

\author{Hui-Feng Fu}
\email{huifengfu@tsinghua.edu.cn} \affiliation{ Department of Physics, Tsinghua University, Beijing 100084, China;}

\author{Guo-Li Wang}
\email{gl_wang@hit.edu.cn} \affiliation{Department of Physics,
Harbin Institute of Technology, Harbin 150001, China;}

\author{Jin-Mei Zhang}
\email{jinmeizhang@tom.com} \affiliation{ Xiamen Institute of Standardization, Xiamen 361004, China.}

\date{\today}

\begin{abstract}
The semileptonic decays $B_c^+\rightarrow P(V) +\ell^++\bar{\nu}_\ell$ and the nonleptonic decays
$B_c^+\rightarrow P(V)+L$, where $P(V)$ denotes a pseudoscalar (vector) charmonium or ($\bar{b}s$)-meson, and $L$ denotes a light meson, are studied in the framework of improved instantaneous
Bethe-Salpeter (BS) equation and the Mandelstam formula. The numerical results (width and branching ratio of the decays) are presented in tables, and in order to compare conveniently, those obtained by other approaches are also put in the relevant tables. Based on the fact that the ratio $\frac{\mathcal{BR}(B_c^+\rightarrow\psi(2S)\pi^+)}{\mathcal{BR}(B_c^+\rightarrow
J/\psi \pi^+)}=0.24^{+0.023}_{-0.040}$ estimated here is in good agreement with the observation by the LHCb $\frac{\mathcal{BR}(B_c^+\rightarrow \psi(2S)\pi^+)}{\mathcal{BR}(B_c^+\rightarrow J/\psi \pi^+)}=0.250\pm0.068(\mathrm{stat})\pm0.014(\mathrm{syst})\pm0.006(\mathcal{B})$, one may conclude
that with respect to the decays the present framework works quite well.
\end{abstract}

\pacs{13.20.He, 13.20.Fc, 13.25.Ft, 13.25.Hw, 11.10.St}

\maketitle

$B_c$ meson carries two heavy flavor quantum numbers explicitly, and
it decays only via weak interactions, although the strong and electromagnetic
interactions can affect the decays. As consequences, $B_c$ meson has a
comparatively long lifetime and very rich weak decay channels with sizable
branching ratios. Being an explicit double heavy flavor meson,
its production cross section can be estimated by perturbative QCD quite reliably
and one can conclude that only via strong interaction and
at hadronic high energy collisions the meson can be produced so
numerously that it can be observed experimentally \cite{07,06,D0}. Therefore, the
meson is specially interesting in studying its production and
decays both.

The first successful observation of $B_c$ was achieved through the
semileptonic decay channel $B_c\rightarrow
J/\psi+\ell^++\bar{\nu}_\ell$ by CDF collaboration
in 1998 from Run-I at Tevatron. They obtained the mass of $B_c$:
$m_{B_c}=6.40\pm0.39\pm0.13$ GeV and the lifetime:
$\tau_{B_c}=0.46^{+0.18}_{-0.16}\pm0.03$ ps \cite{CDF}. Later on CDF
collaboration further gave a more precise mass
$m_{B_c}=6275.6\pm2.9$(stat)$\pm$5(syst) MeV/$c^2$ obtained through
the exclusive non-leptonic decay $B_c\rightarrow J/\psi\pi^+$
\cite{CDF2006}, and upgraded their
results \cite{CDF2008}. In the meantime D0 collaboration at Tevatron also carried
out the observations and confirmed CDF results \cite{D02008}.
Resently, LHCb reported several new observations on $B_c$ decays \cite{LHCb}. Thus we may reasonably expect that at LHCb in the near future the $B_c$ data will be largely enhanced and new results are issued in time.

In literatures, there are many works studying various $B_c$ decays
\cite{Chang,wise,d2,d5,d6,Colangelo,Abd,Anisimov,Ebert,Ivanov,Sanchis,Nobes,Yang,Lusignoli,Liu,Du,Hern,caid,Choi,Xiaozj}
under different approaches.  Among the approaches in the market, the one used in Ref. \cite{Chang}
is that when the components in the concerned meson(s) in initial and final states are heavy quarks, an instantaneous Bethe-Salpeter (BS) equation \cite{BE} (also called Salpeter equation
\cite{E}) with an instantaneous QCD-inspired kernel (interaction)\footnote{With the equation, the spectrum and relevant wave function as an eigenvalue
problem derived from the BS equation can be computed.} is used to depict the meson(s) and the Mandelstam formula \cite{mand} is adopted to compute hadron matrix elements relevant to the concerned decays. This
approach has a comparatively solid foundation because the relativistic `recoil effects'
in the decays\footnote{The difference between masses of the initial $B_c$ meson and the decay product e.g. charmonium is great, so the recoil in a $B_c$ decay must be relativistic, and the "recoil effects" in the decay should be taken into account well.} may be taken into account better than in potential model and else approaches. The reason is that the BS equation and the Mandelstam formula
both are established on relativistic quantum field theory, although the BS equation is deduced into an instantaneous one. Generally, when solving the instantaneous BS equation, the wave function needs to be formulated by a basis of angular momentum with the spin of its components according to the bound state quantum numbers, such as pseudoscalar or vector or else, whereas in Ref. \cite{Chang} to do the formulation the authors, followed Ref. \cite{E}, took an extra
approximation. Since now a way to solve the instantaneous BS equation
without the extra approximation is avaliable \cite{changwang}, and a way more properly to treat the relevant transition matrix elements in Mandelstam formulation has been explored for years \cite{changwang1}, so we think that now it is the right time by using the new wave functions obtained by solving the instantaneous BS equation without the extra approximation and the improved formula for the transition matrix elements to estimate the $B_c$ decays theoretically and then to compare the results with the newly experimental data to see how well the new improved approach \cite{changwang,changwang1} works. Considering the progresses in experiments, especially those at LHCb,
in this paper we would like to restrict ourselves to focus lights on the Cabibbo-Kobayashi-Maskawa (CKM) favored $B_c$ decays: the semileptonic ones $B_c^+\rightarrow P (V)+\ell^++\bar{\nu}_\ell$ and the nonleptonic ones $B_c^+
\rightarrow P (V) +\pi(\rho, K, K^*)$ precisely,
where $P(V)$ represents pseudoscalar (vector) charmonium or a $\bar{b}s$ bound state.

The paper is organized as follows. In Sec. I we outline the useful
formulas. In Sec. II we present numerical results for the
semileptonic and nonleptonic decays and compare the results with those obtained by
other approaches. Sec. III is contributed to discussions. We put the relativistic BS equation with covariant
instantaneous approximation, the forms of relativistic wave functions for
pseudoscalar and vector mesons, the formulations of the form factors, and the parameters used to solve
the BS equation into Appendices.

\section{Formulations for $B_c$ semileptonic and nonleptonic  decays}

For the semileptonic decays $B_c^+ \rightarrow
X+\ell^++\bar{\nu}_\ell$ shown in Fig. 1, the $T$-matrix element can
be written as hadronic component and leptonic component:
\begin{eqnarray}
T=\frac{G_F}{\sqrt{2}}V_{ij}\bar{u}_{\nu_\ell}\gamma^{\mu}(1-\gamma_5)
v_\ell\langle X(p^\prime,\epsilon)|J_{\mu}|B_c^+(p)\rangle,
\end{eqnarray}
where $V_{ij}$ is the CKM matrix
element, $J_{\mu}$ is the charged weak current responsible for the
decays, $p$, $p^{\prime}$ are the momenta of the initial state
$B_c^+$ and the final state $X$ respectively, while $\epsilon$ is
the polarization vector when $X$ is a vector particle. The square of the
matrix element, summed and averaged over the spin (unpolarized), is:
\begin{eqnarray}
\sum\limits^{-}|T|^2
&=&\frac{G_F^2}{2}|V_{ij}|^{2}l^{\mu\nu}h_{\mu\nu},
\end{eqnarray}
where the leptonic tensor:
\begin{eqnarray}
l^{\mu\nu}&\equiv&\bar{u}_{\nu_\ell}\gamma^{\mu}(1-\gamma_5)v_\ell
\bar{v}_\ell(1+\gamma_5)\gamma^{\nu}u_{\nu_\ell},
\end{eqnarray}
is easy to compute, and the hadronic tensor is defined by:
\begin{eqnarray}
h_{\mu\nu}\equiv\sum_\epsilon\langle B_c^+(p)|J_\nu^+|X(p^\prime,
\epsilon)\rangle\langle X(p^\prime,\epsilon)|J_\mu|B_c^+(p)\rangle.
\end{eqnarray}
where $J_\mu=V_{\mu}-A_\mu$. The general form of $h_{\mu\nu}$ based
on Lorentz-covariance analysis can be written as:
\begin{eqnarray}\label{eq4}
h_{\mu\nu}&=&-\alpha
g_{\mu\nu}+\beta_{++}(p+p^{\prime})_{\mu}(p+p^{\prime})_{\nu}\nonumber\\
&&+\beta_{+-}(p+p^{\prime})_{\mu}(p-p^{\prime})_{\nu}+\beta_{-+}
(p-p^{\prime})_{\mu}(p+p^{\prime})_{\nu}\nonumber\\
&&+\beta_{--}(p-p^\prime)_{\mu}(p-p^{\prime})_{\nu}\nonumber\\
&&+i\gamma\epsilon_{\mu\nu\rho\sigma}(p+p^{\prime})^{\rho}(p-p^{\prime})^{\sigma}.
\end{eqnarray}

\begin{figure}
\centering
\includegraphics[width=0.25\textwidth]{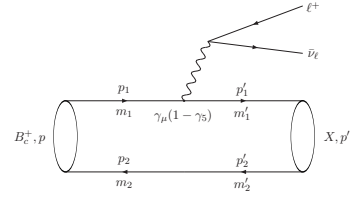}
\caption{\label{fig:epsart}Feynman diagram corresponding to the
semileptonic decays $B_c^+\rightarrow X+\ell^++\bar{\nu}_\ell$.}
\end{figure}

By a straightforward calculation, the differential decay rate is
obtained:
\begin{eqnarray}
\label{diff}
\frac{d^{2}\Gamma}{dxdy}&=&\left|V_{ij}\right|^{2}\frac{G_{F}^{2}
M^{5}}{32\pi^{3}}\left\{\alpha\frac{(y-\frac{m_\ell^2}{M^2})}
{M^2}+2\beta_{++}\right.\nonumber\\
&&\times\left[2x(1-\frac{M^{\prime2}}{M^2}+y)-4x^2-y+\frac{m_\ell^2}
{4M^2}\right.\nonumber\\
&&\left.\times(8x+\frac{4M^{\prime2}-m_\ell^2}{M^2}-3y)\right]\nonumber\\
&&+4(\beta_{+-}+\beta_{-+})\frac{m_\ell^2}{M^2}(2-4x+y
-\frac{2M^{\prime2}-m_\ell^2}{M^2})\nonumber\\
&&+4\beta_{--}\frac{m_\ell^2}{M^2}(y-\frac{m_\ell^2}{M^2})-
\gamma\left[y(1-\frac{M^{\prime2}}{M^2}-4x+y)\right.\nonumber\\
&&\left.\left.+\frac{m_\ell^2}{M^2}(1-\frac{M^{\prime2}}{M^2}+y)\right]\right\},
\end{eqnarray}
where $x\equiv E_\ell/M$ and $y\equiv(p-p^{\prime})^2/M^2$, $M$ is
the mass of $B_c^+$ meson, $M^{\prime}$ is the mass of the final
state $X$. The coefficient functions $\alpha$, $\beta_{++}$, $\beta_{+-}$, $\beta_{-+}$, $\beta_{--}$ and $\gamma$
relate to the form factors of weak currents directly (see below).

To evaluate the exclusive semileptonic differential decay rates of
$B_c^+$ meson, one needs to calculate the hadron matrix element of
the weak current $J_{\mu}$ sandwiched by the $B_c^+$ meson state as
the initial state and a single-hadron state of the concerned final
state, i.e., $\langle
X(p^{\prime},\epsilon)|J_{\mu}|B_c^+(p)\rangle$ with $X$ being a
given suitable meson. In fact, the hadron matrix elements of weak currents can be generally expressed in terms of the momenta $p$ and $p^{\prime}$ of the mesons in initial
state and final state respectivelly, as well as their coefficients. The coefficients, being
functions of the momentum transfer $(p-p^{\prime})$, are
Lorentz-invariant and are called as form factors usually. As emphasised in Refs. \cite{Chang,d6}, with the help of
the Mandelstam formalism \cite{mand}, no matter how great the recoil happens the weak current hadron matrix element can be well calculated, thus here we adopt the method used in Refs.
\cite{Chang,d6} but with improvements \cite{changwang,changwang1}, i.e., the used wave functions are obtained by solving the relevant instantaneous BS equation with the improved approach \cite{changwang,changwang1}. Although the improved approach is used to calculate the hadron matrix elements of weak currents, the form factors are still written as overlap
integrals of the relevant wave functions for the bound states
(mesons). To show the general feature of the improved approach,
we put its outline in Appendices. Moreover, here
we temperately constrain ourselves to consider the cases that $X$ is
a $S$-wave meson only.

According to the Mandelstam formalism and with wave functions of
instantaneous BS equation(s), in the leading order, the matrix
element $\langle X(p^\prime)|J_{\mu}|B_{c}^+(p)\rangle$ can be
written as \cite{changwang1}:
\begin{eqnarray}
\label{eq0001} &&\langle X(p^\prime)|J_\mu|B_{c}^+(p)\rangle\nonumber\\
&=&\int\frac{d^4q^\prime d^4q}{(2\pi)^{4}}Tr\left\{
\bar{\chi}_{p^\prime}J_\mu{\chi}_{p}S^{(2)-1}_{2}(-p_2)
\delta^4(p_2-p^\prime_2)\right\}\nonumber \\
&=&\int\frac{d^3{\vec q}}{(2\pi)^3}Tr\left[
\bar{\varphi}^{++}_{_{p^\prime}}(\vec{q}+\alpha_2^\prime\vec
r)\gamma_{\mu}(1-\gamma_5) {\varphi}^{++}_{_{p}}({\vec
q})\frac{\not\!p}{M}\right].
\end{eqnarray}
here for the last equal sign we have chosen the center of mass
system of initial mason $B_c^+$; $S_2(-p_2)$ is the propagator of
the second component (``spectator"); $\vec r$ is the three
dimensional momentum of finial hadron state $X$ and
$\alpha_2^\prime=m_2^\prime/(m_1^\prime+m_2^\prime)$;
${\varphi}^{++}$ is the component of BS wave function projected onto
the ``positive energy" for the relevant mesons, and may be obtained
by solving the BS equation. Its definition can be found in
Appendix A. Since the initial and finial states in the transition are both heavy
mesons, as adopted in Eq. (\ref{eq0001}), it is a good
approximation that only positive energy projected BS wave functions
are included (the contributions from the component of the wave
functions projected onto the ``negative energy" are much smaller
than that from the positive energy one).

The form factors can be generally related to the
weak current matrix element as follows:

1. If $X$ is a $^1S_0$ state, of the weak current the axial vector matrix element vanishes, and the vector current matrix element can be written as:
\begin{eqnarray}
\langle X(p^{\prime})|V_{\mu}|B_c^+(p)\rangle&\equiv&
f_+(p+p^{\prime})_{\mu}+f_-(p-p^{\prime})_{\mu}.
\end{eqnarray}

2. If $X$ is a $^3S_1$ state, of the weak current the axial vector matrix element can be written as:
\begin{eqnarray}
\langle
X(p^{\prime},\epsilon)|A_{\mu}|B_c^+(p)\rangle&\equiv&f\epsilon_{\mu}^{\ast}
+a_+(\epsilon^{\ast}\cdot p)(p+p^{\prime})_{\mu}\nonumber\\
&&+a_-(\epsilon^{\ast}\cdot p)(p-p^{\prime})_{\mu},
\end{eqnarray}
and the vector current matrix element as:
\begin{eqnarray}
\langle X(p^{\prime},\epsilon)|V_{\mu}|B_c^+(p)\rangle\equiv
ig\epsilon_{\mu\nu\rho\sigma}\epsilon^{\ast\nu}(p+p^{\prime})^{\rho}
(p-p^{\prime})^{\sigma}.
\end{eqnarray}
where $\epsilon$ is the polarization vector of the final hadron $X$.

With the relation between the matrix element and the form factors above and using Eq. (\ref{eq0001}), the
form factors can be calculated out. Explicitly expressions for the form factors as
overlap integrals of meson wave functions are given in Appendix B.3.
Correspondingly, the coefficient functions $\alpha$, $\beta$ and
$\gamma$ in Eq. (\ref{diff}) can be expressed in terms of the form
factors. For example, for the decay $B_c^+\rightarrow
P\ell^+\bar{\nu}_\ell$ ($P$ is a pseudoscalar meson) we have:
\begin{eqnarray}
&\alpha=\gamma=0,\nonumber\\
&\beta_{++}=f_+^2,\beta_{+-}=f_+f_-,
\beta_{-+}=\beta_{+-},\beta_{--}=f_-^2.
\end{eqnarray}
For the decay $B_c^+\rightarrow V\ell^+\bar{\nu}_\ell$ ($V$ is a
vector meson) we have:
\begin{eqnarray}
\alpha&=&f^2+4M^2\vec{p^\prime}^2g^2,\nonumber\\
\beta_{++}&=&\frac{f^2}{4M^{\prime2}}-M^2yg^2+\frac{1}{2}
\left[\frac{M^2}{M^{\prime2}}(1-y)-1\right]fa_+\nonumber\\
&&+M^2\frac{\vec{p^\prime}^2}{M^{\prime2}}a_+^2,\nonumber\\
\beta_{+-}&=&-\frac{f^2}{4M^{\prime2}}+(M^2-M^{\prime2})g^2\nonumber\\
&&+\frac{1}{4}\left[-\frac{M^2}{M^{\prime2}}(1-y)-3\right]fa_+\nonumber\\
&&+\frac{1}{4}\left[\frac{M^2}{M^{\prime2}}(1-y)-1\right]fa_-+
M^2\frac{\vec{p^\prime}^2}{M^{\prime2}}a_+a_-,\nonumber\\
\beta_{-+}&=&\beta_{+-},\nonumber\\
\beta_{--}&=&\frac{f^2}{4M^{\prime2}}+\left[M^2y
-2(M^2+M^{\prime2})\right]g^2\nonumber\\
&&+\frac{1}{2}\left[-\frac{M^2}{M^{\prime2}}(1-y)-
3\right]fa_-+M^2\frac{\vec{p^\prime}^2}{M^{\prime2}}a_-^2,\nonumber\\
\gamma&=&2fg.
\end{eqnarray}

Putting the above form factors into the formula for differential
decay rates Eq. (\ref{diff}), the concerned semileptonic decay rates
can be calculated.

For the nonleptonic decays
$B_c^+ \rightarrow X+\pi(K,\rho,K*)$ concerned here, we follow Ref. \cite{Chang} to
take the CKM-favored effective Hamiltonian with QCD leading logarithm correction
to be responsible for them:
\begin{eqnarray}
&&H^b_{eff}=\frac{G_F}{\sqrt{2}}V_{cb}[c_1^b(\mu_b)Q_1^{cb}+c_2^b(\mu_b)Q_2^{cb}]+h.c.\,,\nonumber\\
&&H^c_{eff}=\frac{G_F}{\sqrt{2}}V_{cs}[c_1^c(\mu_c)Q_1^{cs}+c_2^c(\mu_c)Q_2^{cs}]+h.c.\,
\end{eqnarray}
where $c^c_i(\mu_c)=c^c_i(m_c)$ and $c^b_i(\mu_b)=c^c_i(m_b)$ are the Wilson coefficients,
and the four-fermion operators $Q_1^{ij}$ and $Q_2^{ij}$ are defined:
\begin{eqnarray}
&&Q_1^{bc}\equiv [(\bar{d'}u)_{V-A}+(\bar{s'}c)_{V-A}](\bar{c}b)_{V-A}\,,\nonumber\\
&&Q_2^{bc}\equiv (\bar{c}c)_{V-A}(\bar{s'}b)_{V-A}+(\bar{c}u)_{V-A}(\bar{d'}b)_{V-A}\,,\nonumber\\
&&Q_1^{cs}\equiv (\bar{c}s)_{V-A}(\bar{d'}u)_{V-A}\,,\nonumber\\
&&Q_2^{cs}\equiv (\bar{d'}s)_{V-A}(\bar{c}u)_{V-A}\,,\nonumber
\end{eqnarray}
$d'$ and $s'$ denote 'down' and 'strange' weak eigenstates\footnote{Since we restrict ourselves to consider
the decays $B_c^+ \rightarrow X+\pi(K,\rho,K*)$ here, so we list the main operators the $Q_1^{ij}$ and $Q_2^{ij}$ only
which relate and greatly contribute to the decays.}. Based on the QCD Renormalization Group (RG) calculation,
and in terms of the combination operators $Q_\pm=(Q_1\pm Q_2)$ which have diagonal anomalous dimensions,
the corresponding Wilson coefficients read as follows \cite{Chang,Wise}:
\begin{eqnarray}\label{eq040}
&&\displaystyle c_+^c(\mu)=\Bigg[\frac{\alpha_s(M_W)}{\alpha_s(m_b)}\Bigg]^{6/23}
\Bigg[\frac{\alpha_s(m_b)}{\alpha_s(\mu)}\Bigg]^{6/25}\,,\nonumber\\
&&\displaystyle c_-^c(\mu)=
[c_+^c(\mu)]^{-2}\,\,,\nonumber\\
&&\displaystyle c_+^b(\mu)=\Bigg[\frac{\alpha_s(M_W)}{\alpha_s(m_b)}\Bigg]^{6/23}
\Bigg[\frac{\alpha_s(m_b)}{\alpha_s(\mu)}\Bigg]^{-3/25}\,,\nonumber\\
&&\displaystyle c_-^b(\mu)=\Bigg[\frac{\alpha_s(M_W)}{\alpha_s(m_b)}\Bigg]^{-12/23}
\Bigg[\frac{\alpha_s(m_b)}{\alpha_s(\mu)}\Bigg]^{-12/25}\,,
\end{eqnarray}

Then to use "naive factorization" as done in Ref. \cite{Chang}, the $T$-matrix element can
be written as:
\begin{eqnarray}\label{eq041}
T=\frac{G_F}{\sqrt{2}}V_{ij}V_{lk}a_1\langle L(k,\epsilon^\prime)|J^\mu|0\rangle
\langle X(p^\prime,\epsilon)|J_{\mu}|B_c^+(p)\rangle,
\end{eqnarray}
where $V_{ij}, V_{lk}$ are the relevant CKM matrix elements to $L$ and $X$ accordingly,
$L=\pi (K, \rho, K^*)$, $p, p^\prime, k$ are the momenta of $B_c$, $X$ and
$\pi (K, \rho, K^*)$ respectively, and $\epsilon^\prime, \epsilon$
are the polarization vectors for $\rho$ or $K^*$ and $X$ when $X$ is a vector
meson. The parameter
\begin{eqnarray}\label{eq049}
a_1=c_1(\mu)+\xi\,c_2(\mu)\,,\;\;\; \xi=\frac{1}{N_c}
\end{eqnarray}
in Eq. (\ref{eq041}) is attributed to the contribution from the
operators $Q_1$ and that from the Fierz-reordered $Q_2$ with
a suppressed factor $\xi$
to the concerned decays \cite{Chang}.

For the two-body decays $B_c^+\rightarrow X+L^+$ concerned here,
having the $T$ matrix element Eq. (\ref{eq041}), it is straightforward
to calculate the decay widths.

\section{Numerical Results}

The components of the meson $B_c$ are $\bar{b}$ and $c$ quarks, and it
happens that the contributions from each of them to the total decay rate
are comparable in magnitude. Thus the
semileptonic decay modes of $B_c$ meson can be classified into two:
$\bar{b}$-quark decays with the $c$ quark inside the
meson as a spectator, and $c$-quark decays with the $\bar{b}$ quark
as a spectator. The former causes $B_c$
decays into charmonium or $D$-meson pair, while the latter causes
$B_c$ decays into $B_s$ or $B$ mesons. In this paper, we restrict
ourselves to compute $B_c$ decays to charmonium or $B_s$ meson only
because the approach adopted here is good for double heavy mesons.

When calculating the decays under the adopted approach, we need to fix several parameters. In fact, the parameters are fixed by fitting
well-measured experimental data and the established potential model.
The parameters appearing in the potential (the kernel of Salpeter
equation) used in this work are fixed by the spectra of heavy quarkonia
as done in Ref. \cite{changwang1} and outlined in Appendix B.4.
The masses of the ground states are used as inputs, while the masses of
excited states are considered as predictions. According to the fits we obtain
$M_{\eta_c(2S)}=3.576$ GeV and $M_{\psi(2S)}=3.686$ GeV, and to compare with
experimental data $M^{exp}_{\eta_c'}=3.637$ GeV and $M^{exp}_{\psi'}=3.686$ GeV, one may see the fits are quite good.

\begin{table*}
\caption{\label{tab:table1}The decay widths of the exclusive
semileptonic decay modes (in $10^{-15}$ GeV).}
\begin{ruledtabular}
\begin{tabular}{ccccccccccccc}
Mode&Ours&\cite{Chang}&\cite{d2}&\cite{Abd}&\cite{Anisimov}&\cite{Ebert}
&\cite{Ivanov}&\cite{Sanchis}&\cite{Yang}&\cite{Lusignoli}&\cite{Liu}&\cite{Du}\\
\hline
$B_c^+\rightarrow \eta_ce^+\bar{\nu}_e$&$8.02^{+0.36}_{-0.39}$&14.2&11&11.1&13.05&5.9&14&10&4.3&10.6&8.31&6.5\\
$B_c^+\rightarrow B_se^+\bar{\nu}_e$&$19.7^{+2.0}_{-2.1}$&26.6&59&14.3&22.0&12&29&18&11.75&16.4&26.8&11.1\\
\ \ $B_c^+\rightarrow J/\psi e^+\bar{\nu}_e$&$25.2^{+0.7}_{-0.8}$&34.4&28&30.2&26.6&17.7&33&42&16.8&38.5&20.3&21.8\\
$B_c^+\rightarrow B_s^*e^+\bar{\nu}_e$&$39.9^{+0.6}_{-1.3}$&44.0&65&50.4&51.2&25&37&43&32.56&40.9&34.6&43.7\\
$B_c^+\rightarrow \eta_c(2S)e^+\bar{\nu}_e$&$0.969^{+0.075}_{-0.088}$&0.727&0.28&&&0.46&&&&&0.605&\\
$B_c^+\rightarrow \psi(2S) e^+\bar{\nu}_e$&$1.49^{+0.20}_{-0.25}$&1.45&1.36&&&0.44&&&&&0.186&\\
\end{tabular}
\end{ruledtabular}
\end{table*}

The values of the CKM matrix elements adopted in this paper are \
$V_{cb}=0.0406$,\ $V_{cs}=0.9735$,\ $V_{ud}=0.974$ and $V_{us}=0.2252$.
The properties of relevant light mesons appearing in the concerned
nonleptonic decays are served as phenomenological inputs, namely we take
\begin{eqnarray}
&M_\pi =0.140~\mathrm{GeV}\,,\; \;  &f_\pi=0.130~\mathrm{GeV},\notag\\
&M_\rho =0.775~\mathrm{GeV}\,,\; \;&f_\rho=0.205\pm0.009~\mathrm{GeV},\notag\\
&M_K =0.494~\mathrm{GeV}\,,\; \; &f_K=0.156~\mathrm{GeV},\notag\\
&M_{K^*} =0.892~\mathrm{GeV}\,,\; \;&f_{K^*}=0.217\pm0.005~\mathrm{GeV},\notag
\end{eqnarray}
where the masses and the decay constants are taken from PDG \cite{PDG}, except $f_\rho$ and $f_{K^*}$, which are quoted from Ref. \cite{Ball}.

The numerical results of semileptonic decays are presented in
Table \uppercase\expandafter{\romannumeral1}, and here the uncertainties for our results are obtained by varying the model parameters $m_b$, $m_c$, $m_s$, $\lambda$ and $\Lambda_{\mathrm{QCD}}$ by $\pm5\%$. For comparison precisely, the results from
other typical approaches are also listed in the tables.
To see the feature of the decays, we plot the lepton spectrum for the decays $B_c^+ \rightarrow P
(V)+\ell^++\bar{\nu}_\ell$ in Fig. 2 and Fig. 3 respectively.

\begin{figure}
\centering
\includegraphics[width=2.5in]{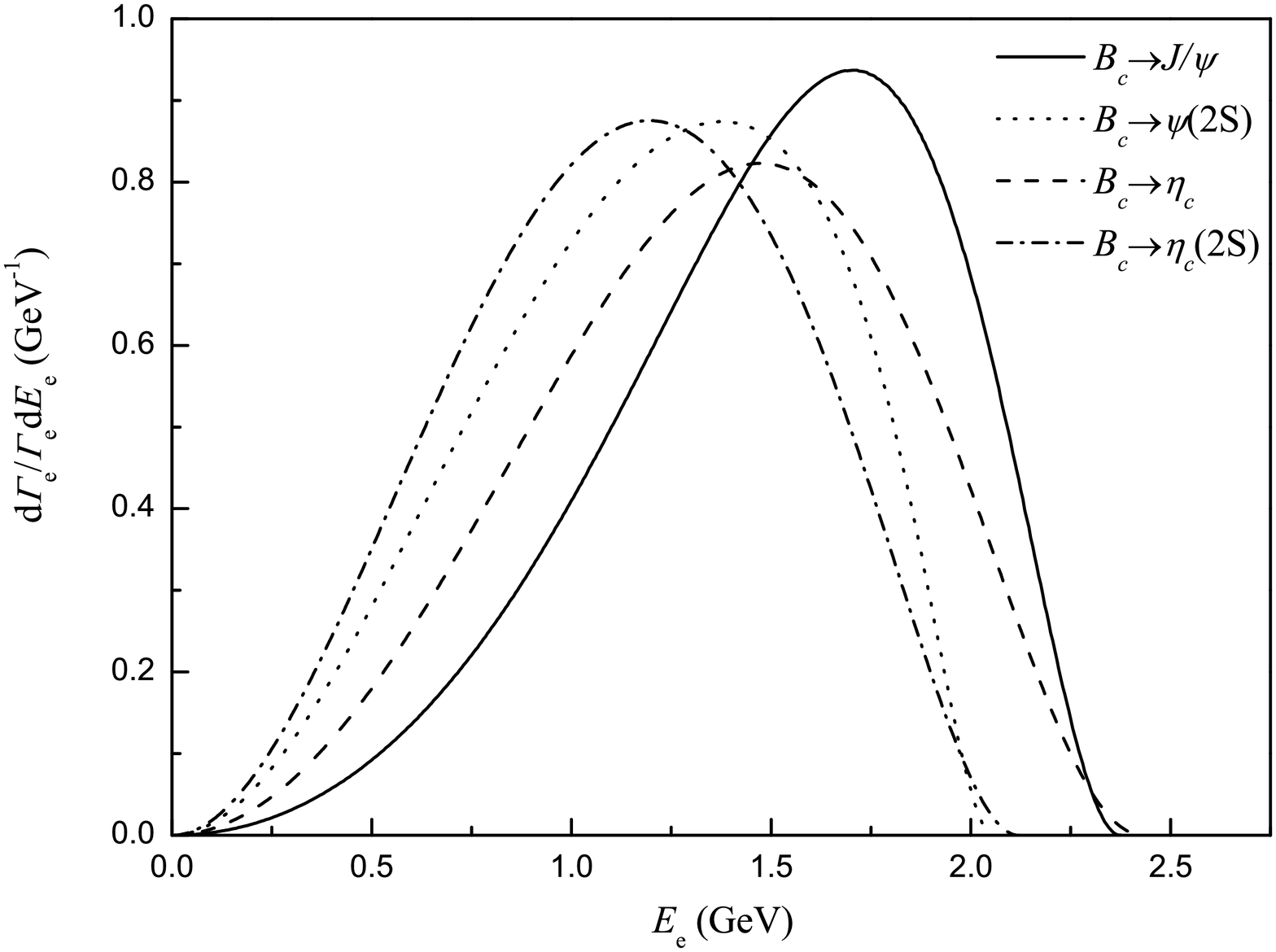}
\caption{\label{fig:epsart1} The lepton energy spectrum for the
semileptonic decays $B_c^+\rightarrow \eta_c(\eta_c(2S), J/\psi, \psi(2S)) \ell^+\bar{\nu}_\ell$.}
\end{figure}

\begin{figure}
\centering
\includegraphics[width=2.5in]{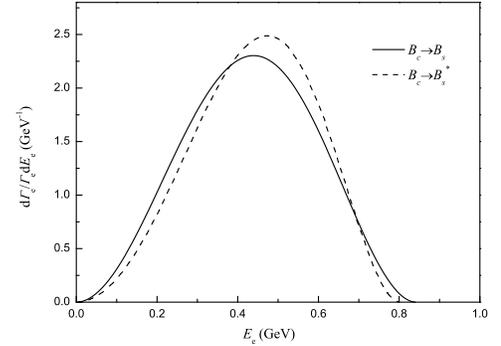}
\caption{\label{fig:epsart2} The lepton energy spectrum for the
semileptonic decays $B_c^+\rightarrow B_s (B_s^*)\ell^+\bar{\nu}_\ell$.}
\end{figure}

The concerned $B_c$ nonleptonic decay modes (some for $\bar b$-decays and $c$ as spectator and some for $c$-decays and $\bar b$ as a spectator) are computed with uncertainties precisely too. The results, as well as some from other approaches for comparisons, are presented
in Table \uppercase\expandafter{\romannumeral2} and Table
\uppercase\expandafter{\romannumeral3}, respectively.

\begin{table*}
\caption{\label{tab:table2}The decay widths of the exclusive
nonleptonic decay modes with $c$-quark spectator (in $10^{-15}$ GeV).}
\begin{ruledtabular}
\begin{tabular}{ccccccccc}
Mode&Ours&\cite{Chang}&\cite{d2}&\cite{Abd}&\cite{Anisimov}&\cite{Ebert}
&\cite{Lusignoli}&\cite{Liu}\\
\hline
$J/\psi+\pi$&$1.24^{+0.11}_{-0.11}a_1^2$&1.97$a_1^2$&1.43$a_1^2$&1.22$a_1^2$&0.82$a_1^2$&0.67$a_1^2$&1.79$a_1^2$&1.01$a_1^2$\\
$J/\psi+K$&$0.0949^{+0.0080}_{-0.0081}a_1^2$&0.152$a_1^2$&0.12$a_1^2$&0.090$a_1^2$&0.079$a_1^2$&0.052$a_1^2$&0.130$a_1^2$&0.0764$a_1^2$\\
$J/\psi+\rho$&$3.59^{+0.64}_{-0.58}a_1^2$&5.95$a_1^2$&4.37$a_1^2$&3.48$a_1^2$&2.32$a_1^2$&1.8$a_1^2$&5.07$a_1^2$&3.25$a_1^2$\\
$J/\psi+K^*$&$0.226^{+0.028}_{-0.029}a_1^2$&0.324$a_1^2$&0.25$a_1^2$&0.197$a_1^2$&0.18$a_1^2$&0.11$a_1^2$&0.263$a_1^2$&0.174$a_1^2$\\
\hline
$\psi(2S)+\pi$&$0.298^{+0.002}_{-0.023}a_1^2$&0.251$a_1^2$&&&&0.12$a_1^2$&&0.0708$a_1^2$\\
$\psi(2S)+K$&$0.0218^{+0.0003}_{-0.0017}a_1^2$&0.018$a_1^2$&&&&0.009$a_1^2$&&0.00499$a_1^2$\\
$\psi(2S)+\rho$&$0.765^{+0.090}_{-0.123}a_1^2$&0.710$a_1^2$&&&&0.20$a_1^2$&&0.183$a_1^2$\\
$\psi(2S)+K^*$&$0.0459^{+0.0037}_{-0.0059}a_1^2$&0.038$a_1^2$&&&&0.011$a_1^2$&&0.00909$a_1^2$\\
\hline
$\eta_c+\pi$&$1.18^{+0.10}_{-0.10}a_1^2$&2.07$a_1^2$&1.8$a_1^2$&1.59$a_1^2$&1.47$a_1^2$&0.93$a_1^2$&1.71$a_1^2$&1.49$a_1^2$\\
$\eta_c+K$&$0.0919^{+0.0078}_{-0.0078}a_1^2$&0.161$a_1^2$&0.15$a_1^2$&0.119$a_1^2$&0.15$a_1^2$&0.073$a_1^2$&0.127$a_1^2$&0.115$a_1^2$\\
$\eta_c+\rho$&$2.89^{+0.51}_{-0.46}a_1^2$&5.48$a_1^2$&4.5$a_1^2$&3.74$a_1^2$&3.35$a_1^2$&2.3$a_1^2$&4.04$a_1^2$&3.93$a_1^2$\\
$\eta_c+K^*$&$0.172^{+0.022}_{-0.021}a_1^2$&0.286$a_1^2$&0.22$a_1^2$&0.200$a_1^2$&0.24$a_1^2$&0.12$a_1^2$&0.203$a_1^2$&0.198$a_1^2$\\
\hline
$\eta_c(2S)+\pi$&$0.322^{+0.010}_{-0.014}a_1^2$&0.268$a_1^2$&&&&0.19$a_1^2$&&0.248$a_1^2$\\
$\eta_c(2S)+K$&$0.0242^{+0.0008}_{-0.0012}a_1^2$&0.020$a_1^2$&&&&0.014$a_1^2$&&0.0184$a_1^2$\\
$\eta_c(2S)+\rho$&$0.711^{+0.094}_{-0.095}a_1^2$&0.622$a_1^2$&&&&0.40$a_1^2$&&0.587$a_1^2$\\
$\eta_c(2S)+K^*$&$0.0408^{+0.0037}_{-0.0040}a_1^2$&0.031$a_1^2$&&&&0.021$a_1^2$&&0.0283$a_1^2$\\
\end{tabular}
\end{ruledtabular}
\end{table*}

\begin{table*}
\caption{\label{tab:table3}The decay widths of the exclusive
nonleptonic decay modes with $b$-quark spectator (in $10^{-15}$ GeV).}
\begin{ruledtabular}
\begin{tabular}{ccccccccc}
Mode&Ours&\cite{Chang}&\cite{d2}&\cite{Abd}&\cite{Anisimov}&\cite{Ebert}
&\cite{Lusignoli}&\cite{Liu}\\
\hline
$B_s+\pi$&$46.5^{+6.2}_{-5.9}a_1^2$&58.4$a_1^2$&167$a_1^2$&15.8$a_1^2$&34.8$a_1^2$&25$a_1^2$&44.0$a_1^2$&65.1$a_1^2$\\
$B_s+K$&$3.55^{+0.38}_{-0.37}a_1^2$&4.20$a_1^2$&10.7$a_1^2$&1.70$a_1^2$&&2.1$a_1^2$&3.28$a_1^2$&4.69$a_1^2$\\
$B_s+\rho$&$26.5^{+4.2}_{-3.9}a_1^2$&44.8$a_1^2$&72.5$a_1^2$&39.2$a_1^2$&23.6$a_1^2$&14$a_1^2$&20.2$a_1^2$&42.7$a_1^2$\\
$B_s+K^*$&$0.0862^{+0.0078}_{-0.0075}a_1^2$&&&1.06$a_1^2$&&0.03$a_1^2$&&0.292$a_1^2$\\
\hline
$B_s^*+\pi$&$31.4^{+2.0}_{-2.4}a_1^2$&51.6$a_1^2$&66.3$a_1^2$&12.5$a_1^2$&19.8$a_1^2$&16$a_1^2$&34.7$a_1^2$&25.3$a_1^2$\\
$B_s^*+K$&$1.66^{+0.06}_{-0.09}a_1^2$&2.96$a_1^2$&3.8$a_1^2$&1.34$a_1^2$&&1.1$a_1^2$&2.52$a_1^2$&1.34$a_1^2$\\
$B_s^*+\rho$&$139^{+11}_{-12}a_1^2$&150$a_1^2$&204$a_1^2$&171$a_1^2$&123$a_1^2$&110$a_1^2$&152.1$a_1^2$&139.6$a_1^2$\\
\end{tabular}
\end{ruledtabular}
\end{table*}

\section{Discussion and Conclusion}

If comparing the semileptonic and nonleptonic decays estimated by various approaches via Tables I-III, one may find that the deviations among the theoretical
predictions by the various approaches are quite wide. Specifically, the results with new
solutions of the Salpeter equation and new formulation are quite different from those
in Ref. \cite{Chang} too.

When calculating the decay branching ratio of semileptonic and nonleptonic decays, here
the lifetime of $B_c$ meson is needed as input. For this purpose, we take the experimental lifetime from PDG \cite{PDG}. For the nonleptonic decays considered here, the parameter $a_1$ for nonleptonic decays appearing in Eq. (\ref{eq041}), additionally,
need to evaluate precisely too. Note that
$a_1$ for $b$ quark (denoted as $a_1^b$) decays should be different from $a_1$ for $c$ quark (denoted as $a_1^c$) decays,
and we take $a_1^b=1.14$ and $a_1^c=1.2$ as in Refs. \cite{d2,Ebert,Ivanov,Hern,Choi}. Having the lifetime and the parameter
$a_1$ fixed, the branching ratio of the concerned decay modes are straightforwardly calculated
and we put the results in Table \uppercase\expandafter{\romannumeral4}
and Table \uppercase\expandafter{\romannumeral5} respectively.

Recently, LHCb has reported an observation of decays $B_c^+\rightarrow \psi\pi^+$ and
$B_c^+\rightarrow \psi(2S)\pi^+$ i.e. the related ratio \cite{LHCb}
\begin{equation}\begin{gathered}
\frac{\mathcal{BR}(B_c^+\rightarrow \psi(2S)\pi^+)}{\mathcal{BR}(B_c^+\rightarrow J/\psi \pi^+)}\\=0.250\pm0.068(\mathrm{stat})\pm0.014(\mathrm{syst})\pm0.006(\mathcal{B}).
\end{gathered}\end{equation}
We would like to point out that, in contrary to the others observables, the above measured ratio, in which the production of $B_c$ meson is canceled totally, is a very essential test of the decays thus here we precisely give the corresponding ratio given by the approach adopted hare:
\begin{equation}
\frac{\mathcal{BR}(B_c^+\rightarrow \psi(2S)\pi^+)}{\mathcal{BR}(B_c^+\rightarrow J/\psi \pi^+)}=0.24^{+0.023}_{-0.040},
\end{equation}
and one may see that it is in good agreement with the observation. Here we should further note that the parameter $a_1$ which appears
in Eq. (\ref{eq041}) and the theoretical uncertainties caused by naive factorization for the
 nonleptonic decays would be canceled a lot in calculating the ratio. Namely the related ratio is mostly determined by hadron transition, so this agreement between the experimental value
and the theoretical estimate on the ratio indicates a vary strong support of the
present approach.

\begin{table}
\caption{\label{tab:table4}The branching ratio (in \%) of the
exclusive semileptonic decay modes with the lifetime of the $B_c$:
$\tau_{B_c}=0.452 ps$.}
\begin{ruledtabular}
\begin{tabular}{lcr}
 Mode&BR (\%)\\\hline
$B_c^+\rightarrow
\eta_ce^+\bar{\nu}_e$&$0.551^{+0.025}_{-0.027}$\\
$B_c^+\rightarrow
B_se^+\bar{\nu}_e$&$1.35^{+0.14}_{-0.14}$\\
$B_c^+\rightarrow J/\psi
e^+\bar{\nu}_e$&$1.73^{+0.05}_{-0.05}$\\
$B_c^+\rightarrow
B_s^*e^+\bar{\nu}_e$&$2.74^{+0.04}_{-0.09}$\\
$B_c^+\rightarrow \eta_c(2S)e^+\bar{\nu}_e$&$0.0665^{+0.0052}_{-0.0060}$\\
$B_c^+\rightarrow \psi(2S) e^+\bar{\nu}_e$&$0.103^{+0.013}_{-0.018}$\\
\end{tabular}
\end{ruledtabular}
\end{table}
\begin{table}
\caption{\label{tab:table5}The branching ratio (in \%) of the
exclusive nonleptonic decay modes with the lifetime of the $B_c$:
$\tau_{B_c}=0.452 ps$.}
\begin{ruledtabular}
\begin{tabular}{cccc}
 Mode&BR (\%)&Mode&BR(\%)\\\hline
$J/\psi+\pi$&$0.111^{+0.009}_{-0.010}$&$\eta_c+\pi$&$0.105^{+0.009}_{-0.009}$\\
$J/\psi+K$&$8.47^{+0.71}_{-0.73}\times 10^{-3}$&$\eta_c+K$&$8.21^{+0.69}_{-0.70}\times 10^{-3}$\\
$J/\psi+\rho$&$0.320^{+0.058}_{-0.051}$&$\eta_c+\rho$&$0.258^{+0.046}_{-0.041}$\\
$J/\psi+K^*$&$0.0201^{+0.0026}_{-0.0025}$&$\eta_c+K^*$&$0.0154^{+0.0019}_{-0.0019}$\\
\hline
$\psi(2S)+\pi$&$0.0266^{+0.0002}_{-0.0020}$&$\eta_c(2S)+\pi$&$0.0287^{+0.0009}_{-0.0012}$\\
$\psi(2S)+K$&$1.94^{+0.03}_{-0.15}\times10^{-3}$&$\eta_c(2S)+K$&$2.16^{+0.07}_{-0.10}\times 10^{-3}$\\
$\psi(2S)+\rho$&$0.0683^{+0.0080}_{-0.0110}$&$\eta_c(2S)+\rho$&$0.0634^{+0.0085}_{-0.0084}$\\
$\psi(2S)+K^*$&$4.10^{+0.32}_{-0.53}\times10^{-3}$&$\eta_c(2S)+K^*$&$3.64^{+0.34}_{-0.36}\times10^{-3}$\\
\hline
$B_s+\pi$&$4.60^{+0.61}_{-0.59}$&$B_s^*+\pi$&$3.11^{+0.19}_{-0.24}$\\
$B_s+K$&$0.352^{+0.036}_{-0.037}$&$B_s^*+K$&$0.164^{+0.006}_{-0.009}$\\
$B_s+\rho$&$2.62^{+0.42}_{-0.39}$&$B_s^*+\rho$&$13.8^{+1.0}_{-1.2}$\\
$B_s+K^*$&$8.53^{+0.77}_{-0.74}\times10^{-3}$&&\\
\end{tabular}
\end{ruledtabular}
\end{table}

In summary, we have calculated the decay width and branching ratio of the exclusive
semileptonic decays of $B_c$ meson to a charmonium or a $B_s$ meson plus leptons
and nonleptonic decays to a charmonium or a $B_s$ meson plus a light meson under the
improved instantaneous BS equation and  Mandelstam approach. Under this approach, the
full Salpeter equations for $(\bar{b}c)$, $(\bar{b}s)$ and $(\bar{c}c)$ etc systems are
solved with the respective full relativistic wave functions for $J^P=0^-$ and
$J^P=1^-$ states. To calculate the hadron transition matrix elements, the Mandelstam formula
has been used and it is suitably approximated to fit the instantaneous approximation. We
find that the results with this approach seem in certain degree to have been improved in comparison with those
obtained by the early ones in Ref. \cite{Chang} with the more approximated formulation. We also should point out that since only the experimental related ratio
$\frac{\mathcal{BR}(B_c^+\rightarrow \psi(2S)\pi^+)}{\mathcal{BR}(B_c^+\rightarrow J/\psi \pi^+)}$
is available now, and the two involved decay modes in the ratio are two-body decays, so the test of the approaches
are limited. Thus we think that to conclude about all the approaches in literature, more experimental data of the semileptonic decays, e.g. the decay spectrum of the positron, and more related ratios of various nonleptonic decays
which are independent on the production of $B_c$ meson etc are requested.

\vspace{4mm}

\noindent {\bf\Large Acknowledgments:}This work was supported
by Nature Science Foundation of China (Grants Nos.
11175051, 11275243 and 11147001), and the Fundamental Research
Funds for the Central Universities and partially Program for Innovation Research
of Science in Harbin Institute of Technology.  We are grateful to Jibo He for indicating the typos and useful discussions.

\appendix
\section{INSTANTANEOUS BS EQUATION}

BS equation for a quark-antiquark bound state generally is written
as:
\begin{eqnarray}
\chi_{_P}(q)=\frac{1}{{\not\!p}_1-m_1}i\int\frac{d^4k}{(2\pi)^4}
V(P,k,q)\chi_{_P}(k)\frac{1}{{\not\!p}_2+m_2},
\end{eqnarray}
where $p_1$, $p_2$; $m_1$, $m_2$ are the momenta and masses of the
quark and anti-quark, respectively. $\chi_{_P}(q)$ is the BS wave
function with the total momentum $P$ and relative momentum $q$,
$V(P,k,q)$ is the kernel between the quark-antiquark in the bound
state. $P$ and $q$ are defined as:
\begin{eqnarray*}
p_1=\alpha_1P+q,\ \alpha_1=\frac{m_1}{m_1+m_2},\\
p_2=\alpha_2P-q,\ \alpha_2=\frac{m_2}{m_1+m_2}.
\end{eqnarray*}
Moreover, the BS wave function $\chi_{_P}(q)$ satisfies the
normalization condition:
\begin{eqnarray}\label{a011}
&\displaystyle
\int\frac{d^4kd^4q}{(2\pi)^4}Tr\left\{\bar{\chi}_{_P}(k)\frac{\partial}
{\partial P_0}\left[S_1^{-1}(p_1)S_2^{-1}(p_2)\delta^4(k-q)\right.\right.\nonumber\\
&\displaystyle
\left.\left.+V(P,k,q)\right]\chi_{_P}(q)\right\}=2iP_0,
\end{eqnarray}
where $S_1(p_1)$ and $S_2(p_2)$ are the propagators of the quark and
anti-quark, respectively.

In general, the BS equation in four dimensional `relative'
space-time is hard to solve comparatively. Whereas if the bound
states are formed by heavy components (quarks) then the kernel of
the equation may approximately become an instantaneous one, and one
may overcome the difficulty to solve the equation in four
dimensional `relative' space-time instead by adopting a so-called
instantaneous approximate approach to turn the equation into a one
in three `relative' space. The proposal by Salpeter \cite{E} is the
approach, that the time-like component of the relative momentum is
integrated out in terms of a contour integration so the BS equation
in four dimension is reduced to a one in three dimension finally
when the kernel is an instantaneous one. For double heavy bound
states, here we follow the Salpeter approach but less approximations
than he did. Let us outline our approach (Salpeter's with less
approximations) here. The approximately instantaneous kernel has the
following form:
\begin{eqnarray}
\label{a00}V(P,k,q)\sim V(|\bf k-\bf q|),
\end{eqnarray}
especially, it is the case, when the two constituents of meson is
very heavy.

Since the recoil in momentum may be great for the concerned $B_c$
semileptonic decays, so for convenience even under instantaneous
approximation we reduce and solve the BS equation in a Lorentz
covariant form, i.e., to divide the relative momentum $q$ into two
parts, $q_{_{P_\|}}$ and $q_{_{P_\bot}}$, a parallel part and an
orthogonal one to $P$, respectively:
\begin{eqnarray}
q^{\mu}=q_{_{P_\|}}^{\mu}+q_{_{P_\bot}}^{\mu},
\end{eqnarray}
where $q_{_{P_\|}}^{\mu}\equiv(P\cdot q/M^2)P^{\mu}$,\
$q_{_{P_\bot}}^{\mu}\equiv q^{\mu}-q_{_{P_\|}}^{\mu}$, and $M$ is
the mass of the relevant meson. Correspondingly, we have two Lorentz
invariant variables:
\begin{eqnarray}
q_{_P}=\frac{P\cdot q}{M},\
q_{_{P_T}}=\sqrt{q_{_P}^2-q^2}=\sqrt{-q^2_{_{P_{\bot}}}}.
\end{eqnarray}

It is easy to see that they turn to the usual component $q_0$ and
$|\vec{q}|$ if in the frame of $\vec{P}=0$. In the same sense, the
volume element of a relative momentum $k$ can be written in an
invariant form:
\begin{eqnarray}
d^4k=dk_{_P}k_{_{P_T}}^2dk_{_{P_T}}dsd\phi,
\end{eqnarray}
where $\phi$ is the azimuthal angle, $s=(k_{_P}q_{_P}-k\cdot
q)/(k_{_{P_T}}q_{_{P_{T}}})$. So now the instantaneous interaction
kernel Eq. (\ref{a00}) can be rewritten as:
\begin{eqnarray}
V(|\vec{k}-\vec{q}|)=V(k_{_{P_{\bot}}},s,q_{_{P_\bot}}).
\end{eqnarray}

If we introduce two notations as below:
\begin{eqnarray}
\eta(q_{_{P_\bot}}^{\mu})&\equiv&\int\frac{k_{_{P_T}}^2dk_{_{P_T}}ds}{(2\pi)^2}
V(k_{_{P_{\bot}}},s,q_{_{P_\bot}})\varphi_{_P}(k_{_{P_{\bot}}}^{\mu}),\nonumber\\
\varphi_{_P}(q_{_{P_{\bot}}}^{\mu})&\equiv&
i\int\frac{dq_{_P}}{2\pi}\chi_{_P}(q_{_{P_\|}}^{\mu},q_{_{P_\bot}}^{\mu}).
\end{eqnarray}

Then the BS equation can be take the form as follows:
\begin{eqnarray}\label{a01}
\chi_{_P}(q_{_{P_\|}}^{\mu},q_{_{P_\bot}}^{\mu})=S_1(p_1^{\mu})
\eta(q_{_{P_\bot}}^{\mu})S_2(p_2^{\mu}).
\end{eqnarray}

The propagator of the relevant particles with masses $m_1$ and $m_2$
can be decomposed as:
\begin{eqnarray}
S_i(p_i^{\mu})&=&\frac{\Lambda_{i{_P}}^+(q_{_{P_\bot}}^{\mu})}{J(i)q_{_P}
+\alpha_iM-\omega_{i{_P}}+i\varepsilon}\nonumber\\
&&+\frac{\Lambda_{i{_P}}^-(q_{_{P_\bot}}^{\mu})}{J(i)q_{_P}+\alpha_iM
+\omega_{i{_P}}-i\varepsilon},
\end{eqnarray}
with
\begin{eqnarray}
\omega_{i{_P}}&=&\sqrt{m_i^2+q_{_{P_T}}^2},\nonumber\\
\Lambda_{i{_P}}^{\pm}(q_{_{P_\bot}}^{\mu})&=&\frac{1}{2\omega_{i{_P}}}
\left[\frac{{\not\!P}}{M}\omega_{i{_P}}\pm
J(i)(m_i+{\not\!q}_{_{P_\bot}})\right],
\end{eqnarray}
where $i$=1, 2 for the quark and anti-quark, respectively, and
$J(i)=(-1)^{i+1}$. $\Lambda_{i{_P}}^{\pm}(q_{_{P_\bot}}^{\mu})$
satisfies the relations as follows:
\begin{eqnarray}
\Lambda_{i{_P}}^+(q_{_{P_\bot}}^{\mu})+\Lambda_{i{_P}}^-(q_{_{P_\bot}}^{\mu})
&=&\frac{{\not\!P}}{M},\nonumber\\
\Lambda_{i{_P}}^{\pm}(q_{_{P_\bot}}^{\mu})\frac{{\not\!P}}{M}
\Lambda_{i{_P}}^{\pm}(q_{_{P_\bot}}^{\mu})&=&\Lambda_{i{_P}}^{\pm}
(q_{_{P_\bot}}^{\mu}),\nonumber\\
\Lambda_{i{_P}}^{\pm}(q_{_{P_\bot}}^{\mu})\frac{{\not\!P}}{M}
\Lambda_{i{_P}}^{\mp}(q_{_{P_\bot}}^{\mu})&=&0.
\end{eqnarray}
In fact, $\Lambda^{\pm}$ may be considered as ``covariant
energy-projection" operators, i.e., in the rest frame $\vec{P}=0$,
they turn to the energy projection operator.

Introducing notations:
\begin{eqnarray}\label{ab02}
\varphi_{_P}^{\pm\pm}(q_{_{P_\bot}}^{\mu})\equiv\Lambda_{1{_P}}^{\pm}
(q_{_{P_\bot}}^{\mu})\frac{{\not\!P}}{M}\varphi_{_P}(q_{_{P_\bot}}^{\mu})
\frac{{\not\!P}}{M}\Lambda_{2{_P}}^{\pm}(q_{_{P_\bot}}^{\mu}),
\end{eqnarray}
and taking into account $\frac{{\not\!P}}{M}\frac{{\not\!P}}{M}=1$,
we have:
\begin{eqnarray*}
\varphi_{_P}(q_{_{P_\bot}}^{\mu})&=&\varphi_{_P}^{++}(q_{_{P_\bot}}^{\mu})
+\varphi_{_P}^{+-}(q_{_{P_\bot}}^{\mu})\nonumber \\
&&+\varphi_{_P}^{-+}(q_{_{P_\bot}}^{\mu})
+\varphi_{_P}^{--}(q_{_{P_\bot}}^{\mu}).
\end{eqnarray*}
Let us further integrate $q_{_P}$ out on both sides of Eq.
(\ref{a01}), and obtain:
\begin{eqnarray*}
\varphi_{_P}(q_{_{P_\bot}}^{\mu})&=&\frac{\Lambda_{1{_P}}^{+}
(q_{_{P_\bot}}^{\mu})\eta_{_P}(q_{_{P_\bot}}^{\mu})\Lambda_{2{_P}}^{+}
(q_{_{P_\bot}}^{\mu})}{M-\omega_{1{_P}}-\omega_{2{_P}}}\nonumber\\
&&-\frac{\Lambda_{1{_P}}^{-}(q_{_{P_\bot}}^{\mu})\eta_{_P}(q_{_{P_\bot}}^{\mu})
\Lambda_{2{_P}}^-(q_{_{P_\bot}}^{\mu})}{M+\omega_{1{_P}}+\omega_{2{_P}}}.
\end{eqnarray*}
We decompose it into the coupled equations:
\begin{eqnarray}\label{a02}
(M-\omega_{1{_P}}-\omega_{2{_P}})\varphi_{_P}^{++}(q_{_{P_\bot}}^{\mu})&=&
\Lambda_{1{_P}}^{+}(q_{_{P_\bot}}^{\mu})\eta_{_P}(q_{_{P_\bot}}^{\mu})
\Lambda_{2{_P}}^{+}(q_{_{P_\bot}}^{\mu}),\nonumber\\
(M+\omega_{1{_P}}+\omega_{2{_P}})\varphi_{_P}^{--}(q_{_{P_\bot}}^{\mu})&=
&-\Lambda_{1{_P}}^{-}(q_{_{P_\bot}}^{\mu})\eta_{_P}(q_{_{P_\bot}}^{\mu})
\Lambda_{2{_P}}^{-}(q_{_{P_\bot}}^{\mu}),\nonumber\\
\varphi_{_P}^{+-}(q_{_{P_\bot}}^{\mu})&=&\varphi_{_P}^{-+}(q_{_{P_\bot}}^{\mu})=\
0.
\end{eqnarray}
Correspondingly, the normalization condition of Eq. (\ref{a011}) in
covariant form reads:
\begin{eqnarray*}
\int\frac{q_{_{P_T}}^2dq_{_{P_T}}}{(2\pi)^2}tr\left[\bar{\varphi}^{++}
\frac{{\not\!P}}{M}\varphi^{++}\frac{{\not\!P}}{M}-
\bar{\varphi}^{--}\frac{{\not\!P}}{M}\varphi^{--}\frac{{\not\!P}}{M}\right]=2P_0.
\end{eqnarray*}

If binding is weak, the positive energy components of the wave
functions $\varphi^{++}$ are large owing to having a very small
factor $(M-\omega_{1{_P}}-\omega_{2{_P}})$, so one can keep the
first equation of Eq. (\ref{a02}) only, and safely dropped the rest
equations at the lowest-order approximation. In Ref. \cite{Chang} it
is the case for the heavy quarkonium and $B_c$ meson.

\section{Precise equation and weak current matrix elements}

The wave functions appearing in the Mandelstam formulas for transition matrix elements are the solution of the corresponding BS equation, so let us show here how to obtain the "precise equation" (all the equations for $\varphi^{\pm\pm}$ are taken into account) and to solve the equation for a concerned heavy meson.

\subsection{\label{sec:level22} Equation and solution for heavy pseudoscalar meson}

The relativistic wave function for heavy pseudoscalar mesons
with the quantum numbers $J^P=0^-$ can be generally written as the
four terms constructed by $P,\ q_{P_\bot}$ and gamma matrices
\cite{d3}:

\begin{eqnarray}\label{aa01}
\varphi^{}_{0^-}(q_{_{P_\bot}})&=&\Big[f_1(q_{_{P_\bot}}){\not\!P}+f_2(q_{_{P_\bot}})M+
f_3(q_{_{P_\bot}})\not\!{q_{_{P_\bot}}}\nonumber\\
&&+f_4(q_{_{P_\bot}})\frac{{\not\!P}\not\!{q_{_{P_\bot}}}}{M}\Big]\gamma_5,
\end{eqnarray}
where $M$ is the mass of the pseudoscalar meson. Due to the last two
equations of Eq. (\ref{a02}):
$\varphi_{0^-}^{+-}=\varphi_{0^-}^{-+}=0$, we have:
\begin{eqnarray}
f_3(q_{_{P_\bot}})&=&\frac{f_2(q_{_{P_\bot}})
M(-\omega_1+\omega_2)}{m_2\omega_1+m_1\omega_2},\nonumber\\
f_4(q_{_{P_\bot}})&=&-\frac{f_1(q_{_{P_\bot}})
M(\omega_1+\omega_2)}{m_2\omega_1+m_1\omega_2}.
\end{eqnarray}
Then there are only two independent wave functions
$f_1(q_{_{P_\bot}})$ and $f_2(q_{_{P_\bot}})$ being left in the Eq.
(\ref{aa01}):
\begin{eqnarray}\label{aa012}
\varphi^{}_{0^-}(q_{_{P_\bot}})&=&\Big[f_1(q_{_{P_\bot}}){\not\!P}
+f_2(q_{_{P_\bot}})M\nonumber\\
&&-f_2(q_{_{P_\bot}}){\not\!{q_{_{P_\bot}}}}
\frac{M(\omega_1-\omega_2)}{m_2\omega_1+m_1\omega_2}\nonumber\\
&&+f_1(q_{_{P_\bot}}){\not\!{q_{_{P_\bot}}}\not\!P}
\frac{\omega_1+\omega_2}{m_2\omega_1
+m_1\omega_2}\Big]\gamma_5.
\end{eqnarray}

According to Eq. (\ref{ab02}) we can further obtain the wave
function corresponding to the positive projection:
\begin{eqnarray}\label{ab03}
\varphi_{0^-}^{++}(q_{_{P_\bot}})&=&L(N+\frac{{\not\!P}}{M}
+\not\!q_{_{P_\bot}}Y+\not\!q_{_{P_\bot}}\frac{{\not\!P}}{M}Z)\gamma_5,
\end{eqnarray}
where
\begin{eqnarray*}
L&=&\frac{M}{2}(f_1+f_2\frac{m_1+m_2}{\omega_1+\omega_2}),\nonumber\\
N&=&\frac{\omega_1+\omega_2}{m_1+m_2},\nonumber\\
Y&=&\frac{m_2-m_1}{m_2\omega_1+m_1\omega_2},\nonumber\\
Z&=&\frac{\omega_1+\omega_2}{m_2\omega_1+m_1\omega_2}.
\end{eqnarray*}

The normalization condition reads:
\begin{eqnarray}
&\displaystyle
\int\frac{d\vec{q}}{(2\pi)^3}4f_1f_2M^2\Big\{\frac{m_1+m_2}{\omega_1
+\omega_2}+\frac{\omega_1+\omega_2}{m_1+m_2}\nonumber\\
&\displaystyle +\frac{2\vec{q}^2(m_1\omega_1+m_2\omega_2)}
{(m_2\omega_1+m_1\omega_2)^2}\Big\}=2M.
\end{eqnarray}

Putting Eq. (\ref{aa012}) into the first two equations of Eq.
(\ref{a02}), we obtain two coupled integral equations about
$f_1(q_{_{P_\bot}})$ and $f_2{_(q_{P_\bot}})$, then by solving them,
we obtain $f_1(q_{_{P_\bot}})$ and $f_2(q_{_{P_\bot}})$, i.e.,
finally the numerical relativistic wave functions Eq. (\ref{aa012})
with $f_1(q_{_{P_\bot}})$ and $f_2(q_{_{P_\bot}})$ being given for
the corresponding pseudoscalar mesons are obtained. Since the $B_c$
and $\eta_c$, $B_s$ etc are pseudoscalar mesons, so the relativistic
wave functions of them, which are needed in calculating the weak
current matrix elements for the concerned semileptonic decays of
$B_c$, are obtained in this way. Note that $s$-quark has a mass
$m_{s}\sim 0.5$ GeV, here we consider it is still ``heavy" although
people consider it is light one, thus for the same reason we are
quite sure that the results about $B_s$ are not so good as those
about $\eta_c$ and $B_c$ etc. The same note for $B^*_s$ is
applicable in the next subsection.

\subsection{\label{sec:level222} Equation and solution for heavy vector meson}

The relativistic wave function of heavy vector state
($J^P=1^-$) generally has 8 terms based on $P,\ q_{_{P_\bot}}$,\
$\epsilon$ (polarization vector) and gamma matrices, so the general
form for the relativistic Salpeter wave function for $1^-$ states
may be read as \cite{changwang1,Wgl}:
\begin{eqnarray}\label{aaa01}
\varphi_{1^-}^\lambda(q_{_{P_\bot}})&=&q_{_{P_\bot}}\cdot
\epsilon_{_\bot}^\lambda\left[f_1(q_{_{P_\bot}})+f_2(q_{_{P_\bot}})
\frac{\not\!P}{M}\right.\nonumber\\
&&\left.+f_3(q_{_{P_\bot}})\frac{\not\!{q_{_{P_\bot}}}}{M}
+f_4(q_{_{P_\bot}})\frac{\not\!P\not\!{q_{P_\bot}}}
{M^2}\right]\nonumber\\
&&+f_5(q_{_{P_\bot}})M\not\!\epsilon_{_\bot}^\lambda
+f_6(q_{_{P_\bot}})\not\!\epsilon_{_\bot}^\lambda\not\!P\nonumber\\
&&+f_7(q_{_{P_\bot}})(\not\!{q_{_{P_\bot}}}{\not\!\epsilon}_{_\bot}^\lambda
-q_{_{P_\bot}}\cdot\epsilon_{_\bot}^\lambda)\nonumber\\
&&+f_8(q_{_{P_\bot}})\frac{(\not\!P\not\!\epsilon_{_\bot}^\lambda\not\!{q_{_{P_\bot}}}
-\not\!Pq_{_{P_\bot}}\cdot\epsilon_{_\bot}^\lambda)}{M},
\end{eqnarray}
where the $M$ is the mass of the vector meson. The equations
$\varphi_{0^-}^{+-}=\varphi_{0^-}^{-+}=0$ give the following
constrains on the components of the wave function:
\begin{eqnarray}
f_1(q_{_{P_\bot}})&=&\left[f_3(q_{_{P_\bot}})q_{_{P_\bot}}^2
+f_5(q_{_{P_\bot}})M^2\right]\nonumber\\
&&\times\frac{(m_1m_2-\omega_1\omega_2
+q_{_{P_\bot}}^2)}{M(m_1+m_2)q_{_{P_\bot}}^2},\nonumber\\
f_7(q_{_{P_\bot}})&=&\frac{f_5(q_{_{P_\bot}})M(-\omega_1+\omega_2)}
{m_2\omega_1+m_1\omega_2},\nonumber\\
f_2(q_{_{P_\bot}})&=&\left[-f_4(q_{_{P_\bot}})q_{_{P_\bot}}^2
+f_6(q_{_{P_\bot}})M^2\right]\nonumber\\
&&\times\frac{(m_1\omega_2-m_2\omega_1)}{M(\omega_1+\omega_2)q_{_{P_\bot}}^2},\nonumber\\
f_8(q_{_{P_\bot}})&=&\frac{f_6(q_{_{P_\bot}})M(\omega_1\omega_2
-m_1m_2-q_{_{P_\bot}}^2)}{(m_1+m_2)q_{_{P_\bot}}^2}.
\end{eqnarray}

Putting the constrains into Eq. (\ref{aaa01}), one can rewrite the
relativistic Salpeter wave function for the states $1^-$ as:
\begin{eqnarray}\label{aaa012}
\varphi_{1^-}^\lambda(q_{_{P_\bot}})&=&q_{_{P_\bot}}\cdot\epsilon_{_\bot}^\lambda
\left\{\frac{\left[f_3(q_{_{P_\bot}})q_{_{P_\bot}}^2
+f_5(q_{_{P_\bot}})M^2\right]}{M(m_1+m_2)}\right.\nonumber\\
&&\times\frac{(m_1m_2-\omega_1\omega_2+q_{_{P_\bot}}^2)}{q_{_{P_\bot}}^2}\nonumber\\
&&+\left[-f_4(q_{_{P_\bot}})q_{_{P_\bot}}^2+f_6(q_{_{P_\bot}})M^2\right]\nonumber\\
&&\times\frac{(m_1\omega_2-m_2\omega_1)\not\!P}{M^2
(\omega_1+\omega_2)q_{_{P_\bot}}^2}\nonumber\\
&&\left.+f_3(q_{_{P_\bot}})\frac{\not\!{q_{_{P_\bot}}}}{M}+f_4(q_{_{P_\bot}})
\frac{\not\!P\not\!{q_{_{P_\bot}}}}{M^2}\right\}\nonumber\\
&&+f_5(q_{_{P_\bot}})M\not\!\epsilon_{_\bot}^\lambda+f_6(q_{_{P_\bot}})
\not\!\epsilon_{_\bot}^\lambda\not\!P\nonumber\\
&&+\frac{f_5(q_{_{P_\bot}})M(-\omega_1+\omega_2)}{m_2\omega_1
+m_1\omega_2}(\not\!{q_{_{P_\bot}}}{\not\!\epsilon}_{_\bot}^\lambda
-q_{_{P_\bot}}\cdot\epsilon_{_\bot}^\lambda)\nonumber\\
&&+\frac{f_6(q_{_{P_\bot}})(\omega_1\omega_2-m_1m_2-q_{_{P_\bot}}^2)}
{(m_1+m_2)q_{_{P_\bot}}^2}\nonumber\\
&&\times(\not\!P\not\!\epsilon_{_\bot}^\lambda\not\!{q_{_{P_\bot}}}
-\not\!Pq_{_{P_\bot}}\cdot\epsilon_{_\bot}^\lambda).
\end{eqnarray}

Furthermore, we can obtain the wave function corresponding to the
positive projection by Eq. (\ref{ab02}):
\begin{eqnarray}\label{abc}
\varphi_{1^-}^{++}(q_{_{P_\bot}})&=&A{\not\!\epsilon}_{_\bot}^\lambda
+B{\not\!\epsilon}_{_\bot}^\lambda\not\!P
+C(\not\!q_{_{P_\bot}}{\not\!\epsilon}_{_\bot}^\lambda
-q_{_{P_\bot}}\cdot\epsilon_{_\bot}^\lambda)\nonumber\\
&&+D(\not\!P\not\!\epsilon_{_\bot}^\lambda\not\!{q_{_{P_\bot}}}
-\not\!Pq_{_{P_\bot}}\cdot\epsilon_{_\bot}^\lambda)
+q_{_{P_\bot}}\cdot\epsilon_{_\bot}^\lambda\nonumber\\
&&\times(E+F\not\!P+G\not\!q_{_{P_\bot}}+H\not\!P\not\!q_{_{P_\bot}}),\nonumber\\
\end{eqnarray}
where
\begin{eqnarray*}
A&=&\frac{1}{2}M(f_5-f_6\frac{\omega_1+\omega_2}{m_1+m_2}),\nonumber\\
B&=&\frac{1}{2}(f_6-f_5\frac{m_1+m_2}{\omega_1+\omega_2}),\nonumber\\
C&=&\frac{1}{2}M\frac{\omega_2-\omega_1}{m_2\omega_1+m_1\omega_2}
(f_5-f_6\frac{\omega_1+\omega_2}{m_1+m_2}),\nonumber\\
D&=&\frac{1}{2}\frac{\omega_1+\omega_2}{\omega_1\omega_2+m_1m_2+\vec{q}^2}
(f_5-f_6\frac{\omega_1+\omega_2}{m_1+m_2}),\nonumber\\
E&=&\frac{1}{2}\frac{m_1+m_2}{M(\omega_1\omega_2+m_1m_2-\vec{q}^2)}
\left[M^2(f_5\right.\nonumber\\
&&\left.-f_6\frac{m_1+m_2}{\omega_1+\omega_2})-\vec{q}^2(
f_3+f_4\frac{m_1+m_2}{\omega_1+\omega_2})\right],\nonumber\\
F&=&\frac{1}{2}\frac{\omega_1-\omega_2}{M^2(\omega_1\omega_2+m_1m_2-\vec{q}^2)}
\left[M^2(f_5\right.\nonumber\\
&&\left.-f_6\frac{m_1+m_2}{\omega_1+\omega_2})-\vec{q}^2
(f_3+f_4\frac{m_1+m_2}{\omega_1+\omega_2})\right],\nonumber\\
G&=&\frac{1}{2}\left[\frac{1}{M}(f_3+f_4\frac{m_1+m_2}{\omega_1+\omega_2})
-\frac{2f_6M}{m_2\omega_1+m_1\omega_2}\right],\nonumber\\
H&=&\frac{1}{2}\frac{1}{M^2}\left[(f_3\frac{\omega_1+\omega_2}
{m_1+m_2}+f_4)-2f_5M^2\right.\nonumber\\
&&\left.\times\frac{\omega_1+\omega_2}{(m_1+m_2)
(\omega_1\omega_2+m_1m_2+\vec{q}^2)}\right].
\end{eqnarray*}

The normalization condition now is read as below:
\begin{eqnarray}
&\displaystyle
\int\frac{d\vec{q}}{(2\pi)^3}\frac{16\omega_1\omega_2}{3}
\left\{3f_5f_6\frac{M^2}{m_2\omega_1+m_1\omega_2}\right.\nonumber\\
&\displaystyle +\frac{\omega_1\omega_2-m_1m_2+\vec{q}^2}
{(m_1+m_2)(\omega_1+\omega_2)}\nonumber\\
&\displaystyle
\times\left.\left[f_4f_5-f_3(f_4\frac{\vec{q}^2}{M^2}+f_6)\right]\right\}=2M.
\end{eqnarray}

From the first two equations of Eq. (\ref{a02}) and in terms of
straightforward calculation, one may obtain four coupled integral
equations about $f_3(q_{_{P_\bot}}),\ f_4(q_{_{P_\bot}}),\
f_5(q_{_{P_\bot}})$ and $f_6(q_{_{P_\bot}})$. By solving them one
may obtain the numerical results for the mass $M$ and the
relativistic wave function Eq. (\ref{aaa012}) with
$f_3(q_{_{P_\bot}}),\ f_4(q_{_{P_\bot}}),\ f_5(q_{_{P_\bot}})$ and
$f_6(q_{_{P_\bot}})$ being given. Since the $J/\psi$ and $B^{*}_s$
etc are vector mesons, so for the concerned semileptonic decays of
$B_c$, all the relativistic wave functions, which are needed in
calculating the weak current matrix elements, are obtained in the
present way.

\subsection{The weak current matrix elements and form factors}

For $B_c^+\rightarrow P\ell^+\bar{\nu}_\ell$ (here we take $P=B_s$
for example), the hadron matrix element Eq. (\ref{eq0001}) based on
the positive energy wave function of pseudoscalar meson Eq.
(\ref{ab03}) can be written as:
\begin{eqnarray}
&&\langle B_s(P^\prime)|J_\mu|B_{c}^+(P)\rangle\nonumber\\
&=&\int\frac{d^3{\vec q}}{(2\pi)^3}4L^\prime L(\frac{P_\mu}{M}s_1
+\frac{P^\prime_\mu}{M^\prime}s_2+q_{_{P_\bot}\mu}s_3)\nonumber \\
&=&S_1\frac{P_\mu}{M}+S_2\frac{P^\prime_\mu}{M^\prime}
+S_3(P^\prime_\mu-\frac{E_f}{M}P_\mu)\nonumber \\
&=&P_\mu(\frac{S_1}{M}-\frac{E_f}{M}S_3)
+P^\prime_\mu(\frac{S_2}{M^\prime}+S_3)\nonumber \\
&=&P_\mu(f_++f_-)+P^\prime_\mu(f_+-f_-)\nonumber \\
&=&f_+(P+P^\prime)_\mu+f_-(P-P^\prime)_\mu,
\end{eqnarray}
where $E_f$ is the energy of the meson in final state,  and
\begin{eqnarray*}
L^\prime&=&\frac{M^\prime}{2}(f_1^\prime+f_2^\prime\frac{m_1^\prime+m_2^\prime}
{\omega_1^\prime+\omega_2^\prime}),\nonumber\\
N^\prime&=&\frac{\omega_1^\prime+\omega_2^\prime}{m_1^\prime+m_2^\prime},\nonumber\\
Y^\prime&=&\frac{m_2^\prime-m_1^\prime}
{m_2^\prime\omega_1^\prime+m_1^\prime\omega_2^\prime},\nonumber\\
Z^\prime&=&\frac{\omega_1^\prime+\omega_2^\prime}
{m_2^\prime\omega_1^\prime+m_1^\prime\omega_2^\prime};
\end{eqnarray*}
\begin{eqnarray*}
s_1&=&N^\prime N+\frac{Y}{M^\prime}\vec r\cdot \vec
q-Y^\prime\alpha_2^\prime E_f\nonumber\\
&&+Y^\prime Y(\vec{q}^2+\alpha_2^\prime \vec r\cdot \vec
q)+\frac{Z^\prime Z}{M^\prime}\alpha_2^\prime E_f\vec r\cdot \vec q\nonumber\\
&&-\frac{Z^\prime N}{M^\prime}(\vec r\cdot \vec q+\alpha_2^\prime
\vec{r}^2+\alpha_2^\prime E_f^2),\nonumber\\
s_2&=&1+Y^\prime\alpha_2^\prime M^\prime+Z^\prime N\alpha_2^\prime
E_f
+Z^\prime Z\vec{q}^2,\nonumber\\
s_3&=&N^\prime Z+\frac{Y}{M^\prime}E_f+Y^\prime +\frac{Z^\prime
N}{M^\prime}E_f\nonumber\\
&&-\frac{Z^\prime Z}{M^\prime}(2\vec r\cdot \vec q+\alpha_2^\prime
\vec{r}^2);
\end{eqnarray*}
\begin{eqnarray*}
S_1&=&\int\frac{d^3{\vec q}}{(2\pi)^3}4L^\prime Ls_1,\nonumber\\
S_2&=&\int\frac{d^3{\vec q}}{(2\pi)^3}4L^\prime Ls_2,\nonumber\\
S_3&=&\frac{1}{|\vec r|}\int\frac{d^3{\vec q}}{(2\pi)^3}|\vec
q|\cos\theta4L^\prime Ls_3;
\end{eqnarray*}
Then the form factors $f_+$ and $f_-$ in Eq. (8) are defined as:
\begin{eqnarray}
f_+&=&\frac{1}{2}(\frac{S_1}{M}+\frac{S_2}{M^\prime}+\frac{M-E_f}{M}S_3),\nonumber\\
f_-&=&\frac{1}{2}(\frac{S_1}{M}-\frac{S_2}{M^\prime}-\frac{M+E_f}{M}S_3).
\end{eqnarray}

For $B_c^+\rightarrow V\ell^+\bar{\nu}_\ell$ (here we take $V=B_s^*$
for example), the hadron matrix element Eq. (\ref{eq0001}) based on
the positive energy wave function of pseudoscalar meson Eq.
(\ref{ab03}) and vector meson Eq. (\ref{abc}) can be written as:
\begin{eqnarray}
&&\langle B_s^*(P^\prime,\epsilon)|J_\mu|B_{c}^+(P)\rangle\nonumber\\
&=&\int\frac{d^3{\vec
q}}{(2\pi)^3}4L\left\{\epsilon_{_\bot\mu}^{\prime\lambda}t_1
+P_\mu\left[(q_{_{P_\bot}}\cdot
\epsilon_{_\bot}^{\prime\lambda})t_2+(P\cdot\epsilon_{_\bot}^{\prime\lambda})
t_2^\prime\right]\right.\nonumber \\
&&+P^\prime_\mu\left[(q_{_{P_\bot}}\cdot
\epsilon_{_\bot}^{\prime\lambda})t_3+(P\cdot\epsilon_{_\bot}^{\prime\lambda})
t_3^\prime\right]\nonumber \\
&&\left.+q_{_{P_\bot}\mu}\left[(q_{_{P_\bot}}\cdot
\epsilon_{_\bot}^{\prime\lambda})t_4+(P\cdot\epsilon_{_\bot}^{\prime\lambda})
t_4^\prime\right]\right.\nonumber \\
&&-i\varepsilon_{\mu\nu\rho\sigma}\left[\frac{A^\prime
Y}{M}\epsilon_{_\bot}^{\prime\lambda\nu}q_{_{P_\bot}}^\rho P^\sigma
-\frac{B^\prime N}{M}P^{\prime\nu}\epsilon_{_\bot}^{\prime\lambda\rho}P^\sigma\right.\nonumber \\
&&-B^\prime
ZP^{\prime\nu}\epsilon_{_\bot}^{\prime\lambda\rho}q_{_{P_\bot}}^\sigma
-\frac{C^\prime N}{M}\epsilon_{_\bot}^{\prime\lambda\nu}q_{_{P_\bot}}^\rho P^\sigma\nonumber \\
&&-\frac{C^\prime
N}{M}\alpha_2^\prime\epsilon_{_\bot}^{\prime\lambda\nu}P^{\prime\rho}
P^\sigma -C^\prime
Z\alpha_2^\prime\epsilon_{_\bot}^{\prime\lambda\nu}P^{\prime\rho}
q_{_{P_\bot}}^\sigma\nonumber \\
&&+\frac{C^\prime Z}{M}\alpha_2^\prime
E_f\epsilon_{_\bot}^{\prime\lambda\nu}P^\rho q_{_{P_\bot}}^\sigma
+D^\prime q_{_{P_\bot}}^\nu\epsilon_{_\bot}^{\prime\lambda\rho}P^{\prime\sigma}\nonumber \\
&&-\frac{D^\prime}{M}\alpha_2^\prime E_f
P^\nu\epsilon_{_\bot}^{\prime\lambda\rho}P^{\prime\sigma}
+\frac{D^\prime Y}{M}\vec{q}^2\epsilon_{_\bot}^{\prime\lambda\nu}P^{\prime\rho}P^\sigma\nonumber \\
&&-\frac{D^\prime Y}{M}\alpha_2^\prime
M^{\prime2}\epsilon_{_\bot}^{\prime\lambda\nu}q_{_{P_\bot}}^{\rho}P^\sigma
-D^\prime Y\alpha_2^\prime E_f\epsilon_{_\bot}^{\prime\lambda\nu}
P^{\prime\rho}q_{_{P_\bot}}^\sigma\nonumber \\
&&-(q_{_{P_\bot}}\cdot
\epsilon_{_\bot}^{\prime\lambda})(\frac{D^\prime
Y}{M}P^{\prime\nu}q_{_{P_\bot}}^{\rho}P^\sigma
-\frac{F^\prime Y}{M}P^{\prime\nu}q_{_{P_\bot}}^{\rho}P^\sigma\nonumber \\
&&-\frac{G^\prime Y}{M}\alpha_2^\prime
P^{\prime\nu}q_{_{P_\bot}}^{\rho}P^\sigma
+\frac{H^\prime N}{M}q_{_{P_\bot}}^\nu P^{\prime\rho}P^\sigma\nonumber \\
&&\left.\left.-\frac{H^\prime Z}{M}\alpha_2^\prime E_fP^\nu
P^{\prime\rho}
q_{_{P_\bot}}^\sigma)\right]\right\}\nonumber \\
&=&(T_1+T_{43})\epsilon_{_\bot\mu}^{\prime\lambda}+(T_2+T_2^\prime+T_{41}+T_{41}^\prime)
(P\cdot\epsilon_{_\bot}^{\prime\lambda})P_\mu\nonumber \\
&&+(T_3+T_3^\prime+T_{42}+T_{42}^\prime)(P\cdot\epsilon_{_\bot}^{\prime\lambda})P^\prime_\mu\nonumber \\
&&+i\varepsilon_{\mu\nu\rho\sigma}\epsilon_{_\bot}^{\prime\lambda\nu}
P^{\prime\rho}P^\sigma(M_1-M_2+M_3+M_4\nonumber \\
&&-M_5+M_6+M_7+M_8-M_9-M_{10}-M_{11}\nonumber \\
&&+M_{12}-M_{13}-V_1+V_2+V_3+V_4)\nonumber \\
&=&f\epsilon_{_\bot\mu}^{\prime\lambda}+
a_+(P\cdot\epsilon_{_\bot}^{\prime\lambda})(P+P^\prime)_\mu\nonumber \\
&&+a_-(P\cdot\epsilon_{_\bot}^{\prime\lambda})(P-P^\prime)_\mu\nonumber \\
&&+ig\varepsilon_{\mu\nu\rho\sigma}\epsilon_{_\bot}^{\prime\lambda\nu}
(P+P^\prime)^\rho(P-P^\prime)^\sigma,
\end{eqnarray}
where the definition of $A^\prime,\ B^\prime,\ C^\prime,\ D^\prime,\
E^\prime,\ F^\prime,\ G^\prime$ and $H^\prime$ is the same as Eq.
(\ref{abc}) but for finial meson, and
\begin{eqnarray*}
t_1&=&A^\prime-B^\prime NE_f+B^\prime Z\vec r\cdot\vec
q\nonumber\\
&&-C^\prime Z(\vec{q}^2+\alpha_2^\prime\vec r\cdot\vec
q)+D^\prime(\alpha_2^\prime\vec{r}^2+\vec r\cdot\vec q)\nonumber\\
&&-D^\prime YE_f(\vec{q}^2+\alpha_2^\prime\vec r\cdot\vec
q),\nonumber\\
t_2&=&-\frac{A^\prime Y}{M}+\frac{D^\prime Y}{M}(\alpha_2^\prime
M^{\prime2}-\vec r\cdot\vec q)-\frac{E^\prime N}{M}\nonumber\\
&&+\frac{F^\prime Y}{M}\vec r\cdot\vec q+\frac{G^\prime
Y}{M}(\vec{q}^2+\alpha_2^\prime\vec r\cdot\vec
q)\nonumber\\
&&-\frac{H^\prime N}{M}(\alpha_2^\prime M^{\prime2}-\vec r\cdot\vec
q)+\frac{C^\prime Z}{M}\alpha_2^\prime
E_f\nonumber\\
&&-\frac{G^\prime}{M}\alpha_2^\prime E_f+\frac{H^\prime
Z}{M}\alpha_2^\prime E_f\vec r\cdot\vec q,
\end{eqnarray*}
\begin{eqnarray*}
t_2^\prime&=&\frac{D^\prime Y}{M}\frac{\alpha_2^\prime E_f}{M}\vec
r\cdot\vec q-\frac{F^\prime Y}{M}\frac{\alpha_2^\prime E_f}{M}\vec
r\cdot\vec
q\nonumber\\
&&-\frac{G^\prime Y}{M}\frac{\alpha_2^\prime
E_f}{M}(\vec{q}^2+\alpha_2^\prime\vec r\cdot\vec
q)\nonumber\\
&&+\frac{H^\prime N}{M}\frac{\alpha_2^\prime E_f}{M}(\alpha_2^\prime
M^{\prime2}-\vec r\cdot\vec
q)\nonumber\\
&&-\frac{H^\prime Z}{M}\frac{\alpha_2^{\prime2} E_f^2}{M}\vec
r\cdot\vec q+\frac{E^\prime
N}{M}\frac{\alpha_2^\prime E_f}{M}\nonumber\\
&&+\frac{C^\prime N}{M}\frac{\alpha_2^\prime
E_f}{M}+\frac{G^\prime}{M}
\frac{\alpha_2^{\prime2} E_f^2}{M},\nonumber\\
t_3&=&B^\prime Z-D^\prime Y\alpha_2^\prime
E_f+F^\prime+H^\prime Z\vec{q}^2\nonumber\\
&&-C^\prime Z\alpha_2^\prime+G^\prime\alpha_2^\prime
+H^\prime N\alpha_2^\prime E_f,\nonumber\\
t_3^\prime&=&\frac{B^\prime N}{M}+\frac{D^\prime Y}{M}\vec{q}^2
-\frac{F^\prime}{M}\alpha_2^\prime E_f
-\frac{C^\prime N}{M}\alpha_2^\prime\nonumber\\
&&-\frac{H^\prime Z}{M}\alpha_2^\prime
E_f\vec{q}^2-\frac{G^\prime}{M}\alpha_2^{\prime2}E_f
-\frac{H^\prime N}{M}\alpha_2^{\prime2}E_f^2,\nonumber\\
t_4&=&-D^\prime YE_f-E^\prime Z+F^\prime YE_f-C^\prime Z\nonumber\\
&&+G^\prime+H^\prime Z\alpha_2^\prime\vec{r}^2+H^\prime NE_f,\nonumber\\
t_4^\prime&=&\frac{A^\prime Y}{M}+\frac{D^\prime
Y}{M}\alpha_2^\prime\vec{r}^2+\frac{E^\prime Z}{M}\alpha_2^\prime
E_f
-\frac{C^\prime N}{M}\nonumber\\
&&-\frac{F^\prime Y}{M}\alpha_2^\prime E_f^2-\frac{H^\prime
Z}{M}\alpha_2^{\prime2} E_f\vec{r}^2\nonumber\\
&&-\frac{G^\prime}{M}\alpha_2^\prime E_f-\frac{H^\prime
N}{M}\alpha_2^\prime E_f^2;
\end{eqnarray*}
\begin{eqnarray*}
T_1&=&\int\frac{d^3{\vec q}}{(2\pi)^3}4At_1,\nonumber\\
T_2&=&-\frac{1}{|\vec r|}\int\frac{d^3{\vec
q}}{(2\pi)^3}\frac{E_f}{M}|\vec
q|\cos\theta4At_2,\nonumber\\
T_2^\prime&=&\int\frac{d^3{\vec q}}{(2\pi)^3}4At_2^\prime,\nonumber\\
T_3&=&-\frac{1}{|\vec r|}\int\frac{d^3{\vec
q}}{(2\pi)^3}\frac{E_f}{M}|\vec
q|\cos\theta4At_3,\nonumber\\
T_3^\prime&=&\int\frac{d^3{\vec q}}{(2\pi)^3}4At_3^\prime,\nonumber\\
T_{41}&=&\frac{1}{2M^2|\vec r|^2}\int\frac{d^3{\vec
q}}{(2\pi)^3}|\vec q|^2\nonumber\\
&&\times\left[(M^{\prime2}+2E_f^2){\cos}^2\theta-M^{\prime2}\right]4At_4,\nonumber\\
T_{41}^\prime&=&-\frac{1}{|\vec r|}\int\frac{d^3{\vec
q}}{(2\pi)^3}\frac{E_f}{M}|\vec
q|\cos\theta4At_4^\prime,\nonumber\\
T_{42}&=&-\frac{E_f}{2M|\vec r|^2}\int\frac{d^3{\vec
q}}{(2\pi)^3}|\vec q|^2(3{\cos}^2\theta-1)4At_4,\nonumber\\
T_{42}^\prime&=&\frac{1}{|\vec r|}\int\frac{d^3{\vec
q}}{(2\pi)^3}|\vec
q|\cos\theta4At_4^\prime,\nonumber\\
T_{43}&=&\frac{1}{2}\int\frac{d^3{\vec
q}}{(2\pi)^3}|\vec q|^2({\cos}^2\theta-1)4At_4;\nonumber\\
\end{eqnarray*}
\begin{eqnarray*}
M_1&=&-\frac{1}{|\vec r|}\int\frac{d^3{\vec q}}{(2\pi)^3}|\vec
q|\cos\theta4L\frac{A^\prime Y}{M},\nonumber\\
M_2&=&-\frac{1}{|\vec r|}\int\frac{d^3{\vec q}}{(2\pi)^3}E_f|\vec
q|\cos\theta4L\frac{B^\prime Z}{M},\nonumber\\
M_3&=&\frac{1}{|\vec r|}\int\frac{d^3{\vec q}}{(2\pi)^3}|\vec
q|\cos\theta4L\frac{C^\prime N}{M},\nonumber\\
M_4&=&-\frac{1}{|\vec r|}\int\frac{d^3{\vec
q}}{(2\pi)^3}\alpha_2^\prime E_f|\vec
q|\cos\theta4L\frac{C^\prime Z}{M},\nonumber\\
M_5&=&-\frac{1}{|\vec r|}\int\frac{d^3{\vec
q}}{(2\pi)^3}\alpha_2^\prime E_f|\vec
q|\cos\theta4L\frac{C^\prime Z}{M},\nonumber\\
M_6&=&\frac{1}{|\vec r|}\int\frac{d^3{\vec q}}{(2\pi)^3}E_f|\vec
q|\cos\theta4L\frac{D^\prime}{M},\nonumber\\
M_7&=&\frac{1}{|\vec r|}\int\frac{d^3{\vec
q}}{(2\pi)^3}\alpha_2^\prime M^{\prime2}|\vec
q|\cos\theta4L\frac{D^\prime Y}{M},\nonumber\\
M_8&=&-\frac{1}{|\vec r|}\int\frac{d^3{\vec
q}}{(2\pi)^3}\alpha_2^\prime E_f^2|\vec
q|\cos\theta4L\frac{D^\prime Y}{M},\nonumber\\
M_9&=&\frac{1}{2}\int\frac{d^3{\vec
q}}{(2\pi)^3}|\vec q|^2({\cos}^2\theta-1)4L\frac{D^\prime Y}{M},\nonumber\\
M_{10}&=&-\frac{1}{2}\int\frac{d^3{\vec
q}}{(2\pi)^3}|\vec q|^2({\cos}^2\theta-1)4L\frac{F^\prime Y}{M},\nonumber\\
M_{11}&=&-\frac{1}{2}\int\frac{d^3{\vec
q}}{(2\pi)^3}\alpha_2^\prime|\vec q|^2({\cos}^2\theta-1)4L\frac{G^\prime Y}{M},\nonumber\\
M_{12}&=&\frac{1}{2}\int\frac{d^3{\vec
q}}{(2\pi)^3}|\vec q|^2({\cos}^2\theta-1)4L\frac{H^\prime N}{M},\nonumber\\
M_{13}&=&-\frac{1}{2}\int\frac{d^3{\vec q}}{(2\pi)^3}\alpha_2^\prime
E_f|\vec q|^2({\cos}^2\theta-1)4L\frac{H^\prime Z}{M};
\end{eqnarray*}
\begin{eqnarray*}
V_1&=&\int\frac{d^3{\vec q}}{(2\pi)^3}4L\frac{B^\prime N}{M},\nonumber\\
V_2&=&\int\frac{d^3{\vec q}}{(2\pi)^3}\alpha_2^\prime4L\frac{C^\prime N}{M},\nonumber\\
V_3&=&\int\frac{d^3{\vec q}}{(2\pi)^3}\alpha_2^\prime E_f4L\frac{D^\prime}{M},\nonumber\\
V_4&=&-\int\frac{d^3{\vec q}}{(2\pi)^3}|\vec q|^24L\frac{D^\prime
Y}{M};
\end{eqnarray*}
Then the form factors $f,\ a_+,\ a_-$ and $g$ in Eq. (9) and (10)
are defined as:
\begin{eqnarray}
f&=&T_1+T_{43},\nonumber\\
a_+&=&\frac{1}{2}(T_2+T_2^\prime+T_{41}+T_{41}^\prime\nonumber\\
&&+T_3+T_3^\prime+T_{42}+T_{42}^\prime),\nonumber\\
a_-&=&\frac{1}{2}(T_2+T_2^\prime+T_{41}+T_{41}^\prime\nonumber\\
&&-T_3-T_3^\prime-T_{42}-T_{42}^\prime),\nonumber\\
g&=&\frac{1}{2}(M_1-M_2+M_3+M_4-M_5\nonumber\\
&&+M_6+M_7+M_8-M_9-M_{10}\nonumber\\
&&-M_{11}+M_{12}-M_{13}-V_1\nonumber\\
&&+V_2+V_3+V_4).
\end{eqnarray}

\subsection{The Parameters in QCD inspired BS Equation}

When solving the equations, we have to fix the BS (instantaneous) kernel.
Considering the successes of Cornell potential model on heavy quarkonia\cite{EK}, we would like to refer the BS kernel to the model. Moreover, the color factor for
the relevant BS equation may be factorized out straightforwardly, thus we leave the
factor aside, and focus on the rest factors of the formulation for the
kernel. They are a linear scalar interaction
$V_s(r)=\lambda r$ for `color-confinement', a vector interaction
$V_\upsilon(r)=-\frac{4}{3}\frac{\alpha_s(r)}{r}$ for one-gluon exchange, i.e.:
\begin{eqnarray}\label{ccc01}
I(r)&=&V_s(r)+V_0+\gamma_0\otimes\gamma^0V_\upsilon(r)\nonumber\\
&=&\lambda r+V_0-\gamma_0\otimes\gamma^0\frac{4}{3}\frac{\alpha_s(r)}{r},
\end{eqnarray}
where $\lambda$ is the the so-called `string constant', $\alpha_s(r)$ is the running
coupling constant, and a constant $V_0$, which, as a `zero-point', is added.

The kernel in momentum space reads:
\begin{eqnarray}
\label{ccc03}
I(\vec{q})&=&V_s(\vec{q})+\gamma_0\otimes\gamma^0V_\upsilon(\vec{q})\,,
\end{eqnarray}
where
$$V_s(\vec{q})=-\Big(\frac{\lambda}{\alpha}+V_0\Big)\delta^3(\vec{q})
+\frac{\lambda}{\pi^2}\frac{1}{(\vec{q}^2+\alpha^2)^2}\,,$$
$$V_\upsilon(\vec{q})=-\frac{2}{3\pi^2}\frac{\alpha_s(\vec{q})}{(\vec{q}^2+\alpha^2)}\,,$$
and $$\alpha_s(\vec{q})=\frac{12\pi}{27}\frac{1}{\log(a+\frac{\vec{q}^2}{\Lambda_{QCD}^2})}$$.

In order to avoid the Coulomb-like infrared divergence,
usually a factor $e^{-\alpha r}$ as below:
\begin{eqnarray}\label{ccc02}
V_s(r)&=&\frac{\lambda}{\alpha}(1-e^{-\alpha r}),\nonumber\\
V_\upsilon(r)&=&-\frac{4}{3}\frac{\alpha_s(r)}{r}e^{-\alpha r}.
\end{eqnarray}
is introduced.

The parameters $\lambda,\ \alpha,\ a$ and $\Lambda_{QCD}$
characterizing the potential are fixed by fitting the mass
spectrum of heavy quarkonium \cite{changwang1}. The fitted values are $a=e=2.7183,$\ $\alpha=0.06$
GeV,\ $\lambda=0.21$ GeV$^2$, $\Lambda_{QCD}=0.27$ GeV. The parameter $V_0$ varies as the constituents
and the quantum number of the concerned meson being varying. In this work, the relevant values are $V_0=-0.314$ GeV for
$\bar{c}c (0^{-+})$, $V_0=-0.176$ GeV for $\bar{c}c (1^{--})$, $V_0=-0.205$ GeV for
$\bar{b}s (0^{-})$, $V_0=-0.13$ GeV for $\bar{b}s (1^{-})$ and $V_0=-0.185$ GeV for
$\bar{b}c (0^{-})$. The constituent quark masses are parameters too, and they
are fixed by fitting the meson spectrum:\ $m_b=4.96$ GeV,\ $m_c=1.62$
GeV,\ $m_s=0.5$ GeV. With these parameters, we obtain the mass spectrum and the relevant wave functions by solving the precise equations obtained in previous subsection.

\end{document}